\documentclass[aps,pre,twocolumn,showpacs,floatfix]{revtex4-1}
\usepackage{graphicx}
\usepackage{amssymb}
\usepackage{color}
\usepackage{amsmath}
\usepackage{ulem}
\usepackage{mathrsfs}

\newcommand{\be}{\begin{equation}}
\newcommand{\ee}{\end{equation}}

\begin{document}

\title{Dual time scales in simulated annealing of a two-dimensional Ising spin glass} 
  
\author{Shanon J. Rubin} 
\affiliation{Department of Physics, Boston University, 590 Commonwealth Avenue, Boston, Massachusetts 02215, USA}

\author{Na Xu}
\email{naxu@buphy.bu.edu}
\affiliation{Department of Physics, Boston University, 590 Commonwealth Avenue, Boston, Massachusetts 02215, USA}

\author{Anders W. Sandvik}
\email{sandvik@buphy.bu.edu}
\affiliation{Department of Physics, Boston University, 590 Commonwealth Avenue, Boston, Massachusetts 02215, USA}

\begin{abstract}
We apply a generalized Kibble-Zurek out-of-equilibrium scaling ansatz to simulated annealing when approaching the spin-glass transition at 
temperature $T=0$ of the two-dimensional Ising model with random $J= \pm 1$ couplings. Analyzing the spin-glass order parameter and the
excess energy as functions of the system size and the annealing velocity in Monte Carlo simulations with Metropolis dynamics, we find scaling
where the energy relaxes slower than the spin-glass order parameter, i.e., there are two different dynamic exponents. The values of the exponents
relating the relaxation time scales to the system length, $\tau \sim L^z$, are
$z=8.28 \pm 0.03$ for the relaxation of the order parameter and $z=10.31 \pm 0.04$ for the energy relaxation. We argue that the behavior with dual
time scales arises as a consequence of the entropy-driven ordering mechanism within droplet theory. We point out that the dynamic exponents found here
for $T \to 0$ simulated annealing are different from the temperature-dependent equilibrium dynamic exponent $z_{\rm eq}(T)$, for which previous studies
have found a divergent behavior: $z_{\rm eq}(T\to 0) \to \infty$. Thus, our study shows that, within Metropolis dynamics, it is easier to relax the
system to one of its degenerate ground states than to migrate at low temperatures between regions of the configuration space surrounding different
ground states. In a more general context of optimization, our study provides an example of robust dense-region solutions for which the excess
energy (the conventional cost function) may not be the best measure of success. 
\end{abstract}

\date{\today}


\maketitle

\section{Introduction}

A simulated annealing (SA) process \cite{kirkpatrick83} carried out on a system with a continuous phase transition 
exhibits scaling with the system size and the annealing velocity (the rate of change of the temperature 
$T$ versus time). Following the seminal analysis by Kibble \cite{kibble76} and Zurek \cite{zurek85} (KZ) of the
``freezing'' of defects close to a critical point, a compelling picture has emerged of the combined effects of finite 
size and velocity on physical observables in SA
\cite{zhong05,chandran12,liu14,liu15}. The generalization of KZ scaling to quantum systems (where a system parameter is 
changed as a function of time at low $T$) \cite{polkovnikov05,zurek05,dziarmaga05} has found applications in studies of 
cold atom systems \cite{lamporesi13,clark16}, and should be of relevance also in the quantum annealing (QA) \cite{kadowaki98} 
(or quantum-adiabatic \cite{farhi01}) approach to solving hard optimization problems by adiabatically evolving a programmable 
qubit system from a trivial to a complex ground state \cite{liu15a}.

An untested application of classical KZ scaling is to systems with critical temperature $T_c=0$. A prominent example of such a 
case is the two-dimensional (2D) Ising spin glass, which is interesting not only in its own right but also in the context of 
quantum annealing, e.g., the devices produced by D-Wave Systems \cite{johnson11} are laid out with a particular 2D connectivity. 
Numerous studies of 2D Ising spin glasses have been carried out recently in order to compare SA and QA, to gain insights into 
the nature of the quantum and thermal fluctuations in QA devices, and to develop methods for analyzing the efficiency of annealing 
protocols \cite{boixo14,katzgraber14,heim15,katzgraber15}. The KZ scaling formalism has not been applied, however. We here present
such a study of the 2D Ising spin glass with bimodal couplings and find an unusual behavior where two different dynamic exponents
govern the equilibration of the spin-glass order parameter and the excess energy in SA simulations with local Metropolis
Monte Carlo (MC) dynamics. The exponent for the
order parameter is $z=8.28 \pm 0.03$ and for the excess energy $z'=10.31 \pm 0.04$, and, thus, the energy relaxes slower than the 
order parameter. We argue that this unusual behavior is a consequence of the entropy-driven spin-glass ordering process within
droplet theory \cite{fisher88,thomas11}, by which the system can first reach the region of high density of low-energy states,
where the order parameter is not sensitive to the energy, and only later relax to the minimum energy.

Our results also show that the dynamics of simulated annealing is not necessarily governed by the dynamic exponent $z_{\rm eq}$ of the
equilibrium  autocorrelation functions, though at conventional critical points at $T>0$ this is the case in all systems we are aware of, e.g., the
standard 2D and 3D Ising models with Metropolis and cluster dynamics \cite{liu14} and the 3D Ising spin glass \cite{liu15}. In the 2D Ising glass
$z_{\rm eq}$ depends on the temperature and diverges as $T \to 0$ \cite{liang92,katzgraber05}, in contrast to the finite values of $z$ and $z'$
found here for $T\to 0$ SA simulations.

In Sec.~\ref{sec:methods} we define the Ising spin-glass model, discuss its known equilibrium properties, and describe the simulation methods
we have used to study it. In addition to SA, we also implemented parallel tempering (PT) for obtaining equilibrium $T \to 0$ results (which are
later used together with SA data in the KZ analysis). We discuss equilibrium  finite-size scaling in Sec.~\ref{sec:fs}. The KZ scaling ansatz and
its connections to both the equilibrium and high-velocity behavior is outlined in Sec.~\ref{sec:kz} and adapted to the particular circumstances
of the $T \to 0$ relaxation of the Ising spin glass. Results are presented in support of the dual time-scale behavior. In Sec.~\ref{sec:disc},
we discuss the physical reasons behind our findings and the significance of dual SA time scales in a more general context of optimization.
Additional analysis and results are presented in two appendices.

\section{Model and methods}
\label{sec:methods}

The 2D Ising spin glass considered here is defined by the Hamiltonian
\begin{equation}
H = \sum_{\langle ij\rangle}J_{ij} \sigma_{i} \sigma_{j},~~~ \sigma_i=\pm 1,
\label{ham}
\end{equation}
with random nearest-neighbor couplings  $J_{ij}$ drawn from some distribution, e.g., bimodal or normal (with the former used here). A central property
of spin glasses is the Edwards-Anderson (EA) order parameter, defined with two replicas (independent simulations, labeled $1$ and $2$, with the same
couplings) as
\begin{equation}
q = \frac{1}{N}\sum_{i=1} \sigma^{(1)}_i\sigma^{(2)}_i.
\label{ea}
\end{equation}
We focus our studies in this paper on the disorder-averaged squared EA order parameter $\langle q^2\rangle$ and the internal energy $E=\langle H\rangle/N$
in the limit $T\to 0$. 

The equilibrium 
properties of the bimodal $J_{ij} = \pm 1$ model have been controversial. A long-standing issue has been to distinguish
between exponential \cite{wang88,campbell04} and power-law \cite{kawashima92,rieger96,katzgraber07} scaling as $T \to 0$. The nature of the state at 
$T=0$ has also been difficult to ascertain. Until recently it was widely believed that the $J_{ij} = \pm 1$ system does not harbor spin-glass 
order (unlike the model with normal-distributed couplings), only power-law decaying critical EA spin-spin correlations 
\cite{morgenstern80,bhatt88,poulter05}. More recent studies \cite{jorg06,roma10,thomas11} point to significant long-range order. In particular,
Thomas {\it et al.} \cite{thomas11} evaluated the Pfaffian form of the partition function on larger lattices and lower temperatures than
in previous MC studies. A quantitative picture was presented for finite-size corrections to the long-range order at $T=0$,
power-law scaling at $T>0$, and a size-dependent cross-over temperature $T^*(L)$, with $T^*\to 0$ when $L \to \infty$, below which the
discreteness of the coupling distribution is important. 

We here use out-of-equilibrium (SA) MC simulations to study the model (\ref{ham}) with $N=L^2$ spins on periodic square lattices with bimodal
coupling distribution. We generate the
$J_{ij}=\pm 1$ couplings independently with probability $1/2$ and use bit representations for both the spins and the couplings, as discussed in detail
in Sec.~\ref{subsec:sa}, running 64 independent parallel simulations for each realization (sample) of the couplings and repeating for a large number
of samples. In addition to SA, where we go to system sizes up to $L=128$, we have also used PT simulations \cite{hukushima96} to obtain equilibrium
$T \to 0$ results for smaller systems (up to $L=32$ for the energy and $L=24$ for the EA order parameter). Technical details and convergence
tests of the PT simulations are presented below in Sec.~\ref{subsec:pt}. Although larger systems (with open boundaries) can be studied with ground-state
methods \cite{campbell04,palmer99}, proper thermodynamic averages of the EA order parameter, with equal weighting of degenerate states, are difficult
to obtain \cite{sandvik99}.

\subsection{Simulated annealing}
\label{subsec:sa}

We code the Ising spins $\sigma_i = \pm 1$ of the model (\ref{ham}) as bits of long (64-bit) integers, thus using $N$ integers $I_i$ 
for a system of $N$ spins and propagating 64 replicas of the same system (with the same random couplings). The bimodal couplings 
$J_{ij} = \pm 1$ are also encoded as bits $0,1$, and most of the operations involved in computing energy differences for the Metropolis 
acceptance probabilities for single-spin flips (with the same spin considered in all replicas) can then be carried out simultaneously 
on all 64 replicas by using standard bit-vise logical operations on the stored integers. 

In the beginning of each repetition of the SA process, we generate new random couplings and initialize the spins at random. 
We then carry out 10 MC sweeps at the initial temperature $T_{\rm ini}=8$. We found that this small number of initial steps is sufficient 
for reaching very close to thermal equilibrium at this high temperature (and note that any deviation from equilibrium at this stage can 
be regarded as just a different initial state and will not affect the scaling when $T \to 0$ at low velocities). In the subsequent 
SA run we carry out $t_{\rm max}$ MC sweeps and lower the temperature after each sweep according to the following generic power-law protocol
to anneal the system to $T=0$:
\begin{equation}
T(t) = T_{\rm ini}(1-t/t_{\rm max})^r.
\label{tprotocol}
\end{equation}
In addition to the linear case $r=1$, we also study $r>1$.
Measurements of the EA order parameter and the energy are carried out after the final ($T=0$) MC sweep and results are averaged over a large
number of SA runs. To compute the EA order parameter (\ref{ea}), we form 32 configuration pairs out of the 64 replicas and again make use of bit
operations for parallel computing, thus obtaining 32 independent contributions to $\langle q^2\rangle$ from each run.

The safest way to ensure independent propagation of the replicas is to generate different random numbers for the final Meropolis accept-reject
step for each replica, in which case the generation of the random numbers consumes a large fraction of the computation time. Strictly speaking,
uncorrelated replicas are required only when computing the EA order parameter; correlations of the replicas do not cause distortions of computed
averages (provided that the random number generator is not flawed), though the efficiency is potentially reduced as there is effectively a smaller
amount of statistical data. For example, if the same random number is used for each replica, if ever two replicas go into the same state they
will stay in the same state for the remainder of the simulation, thus reducing the number of independent replicas. No statistical bias is
introduced in computed mean values, however. Once the system size is reasonably large, it is very unlikely for replicas to lock to each other
in this way, and we can safely use the same random numbers within the two groups of 32 replicas between which the EA order is computed.

\subsection{Parallel tempering}
\label{subsec:pt}

In our PT simulations \cite{hukushima96}, we again use the bit representation, but now all the bits $b \in \{0,63\}$ correspond
to different temperatures on a uniform grid, $T_b=T_0 + b\Delta_T$. Attempts to swap spin configurations of runs at adjacent temperatures 
$T_b,T_{b+1}$ are carried out after each MC sweep over the spins, with independent random numbers used for the MC updates at all
temperatures. The goal of the PT simulations is to obtain $T \to 0$ equilibrium results for the EA order parameter and the ground-state energy.
For the latter, we do not use the thermal energy average but simply keep track of the lowest energy reached in each run and average it over
the coupling samples. We here present results showing proper convergence to equilibrium values of computed
quantities as the number of MC sweeps is increased.

We choose the lowest temperature $T_0$ such that the $T$ dependence of the energy and the EA order parameter is insignificant in the neighborhood of 
this temperature for the system sizes studied, i.e., $T$ is well below the size-dependent entropic cross-over temperature $T^*(L)$ \cite{thomas11}
mentioned above and discussed in detail in Sec.~\ref{sec:kz}. The highest temperature should be high enough for significant thermal fluctuations to
migrate to low temperatures, thereby enhancing the ergodicity of the PT simulations relative to independent fixed-$T$ runs. Efficient migration of 
the fluctuations also necessitates a sufficiently small spacing $\Delta_T$, and, in principle optimal simulations would have $\Delta_T$ decreasing and 
the number of temperatures increasing with the system size. Here we always use $64$ replicas and the spacing is $\Delta_T=0.04$ or $0.02$, for
smaller and larger lattices, respectively.

\begin{figure}[tp]
\center{\includegraphics[width=7.25cm, clip]{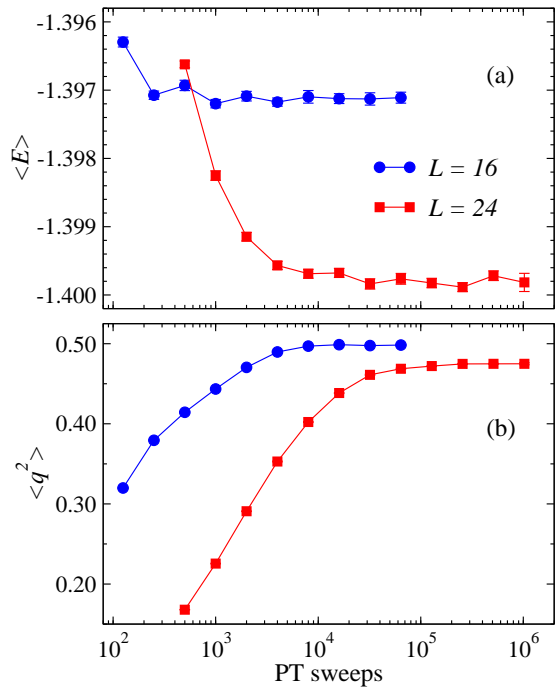}}
\vskip-2mm
\caption{Convergence of PT results of the EA order parameter (a) and the excess energy (b) vs the number of
  sweeps for two system sizes. The lowest temperature was $T_0=0.1$ and $T_0=0.06$ for $L=16$ and $24$, respectively, and in both
  cases the temperature spacing was $\Delta_T=0.02$. The results represent averages over more than $10^5$ samples for both system size.}
\label{ptconv}
\vskip-2mm
\end{figure}

\begin{figure}
\center{\includegraphics[width=7.75cm, clip]{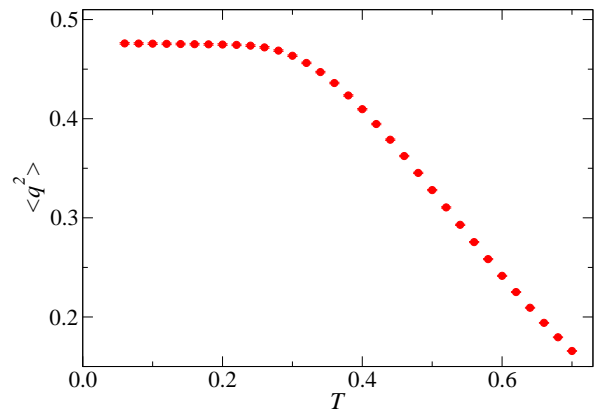}}
\vskip-2mm
\caption{Temperature dependence of the squared EA order parameter for $L=24$ in PT simulations with $10^6$ sweeps for both equilibration
  and data collection.}
\label{m24}
\vskip-2mm
\end{figure}

Figure \ref{ptconv} shows examples of the convergence of the EA order parameter and the lowest energy reached as functions of the number of MC
sweeps in PT simulations. We use the same number of MC sweeps for equilibration and data collection. The horizontal axis of Fig.~\ref{ptconv}
corresponds to the sweeps for data collection only (i.e., the total number of sweeps is twice this number) and each successive point
corresponds to doubling the number of sweeps. For these system sizes, $L=16,24$, the energy converges faster than the order
parameter, but this trend is clearer for $L=16$ than $L=24$. The energy likely converges {\it slower} than the order parameter
for large sizes, as we find in SA simulations in Sec.~\ref{sec:kz}. We have not studied the scaling properties of the PT scheme
in detail.

For acceptable convergence, we require statistically indistinguishable results from at least the last two runs in a series of runs such as those
in Fig.~\ref{ptconv}. Based on this criterion we have obtained converged results for $\langle q^2\rangle$ up to $L=24$ and for $\langle E\rangle$
up to $L=32$. To ensure that we obtain $T\to 0$ results, it is also important to check the temperature dependence of the results. Fig.~\ref{m24}
shows results for $L=24$ from the PT runs with the largest number of sweeps in Fig.~\ref{ptconv}(b). Here we can see that there is only a weak
temperature dependence below $T\approx 0.25$. We estimate that the very small remaining finite-temperature effect at $T_0=0.06$ is much smaller
than the statistical error.

\section{Equilibrium finite-size scaling}
\label{sec:fs}
        
We discuss the equilibrium PT results first because they will be important in the KZ scaling analysis. The mean values presented here were
computed over millions of coupling samples for the smaller system sizes and about $10^5$ samples for the largest systems.

\begin{figure}
\center{\includegraphics[width=7cm, clip]{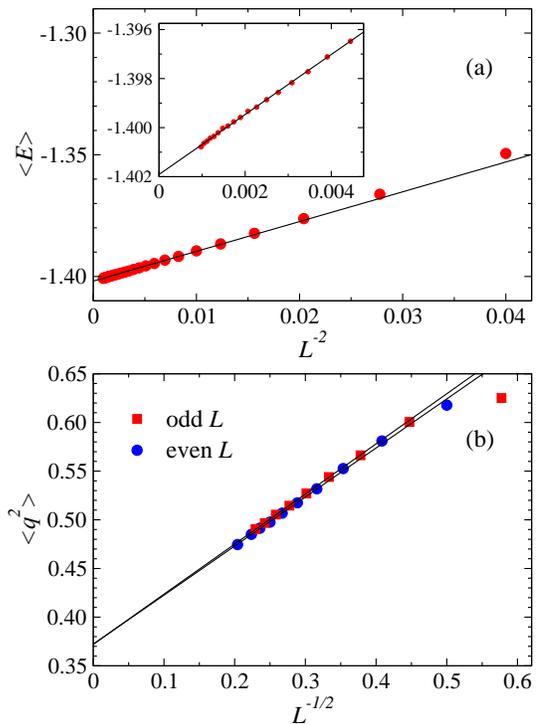}}
\vskip-2mm
\caption{Equilibrium results for (a) the ground state energy graphed vs $L^{-2}$ and (b) the EA order parameter vs $L^{-1/2}$. In (a) the
  results for $L=14-32$ were fitted to $E_0=-1.401938$ \cite{palmer99} plus a correction $a L^{-2}$ with $a=1.230(2)$. The inset shows
  data for the larger systems on a more detailed scale (where the error bars are similar to the symbol size). In (b), there are 
  significant even-odd effects and $L\ge 9$ data for even (blue circles) and odd (red squares) $L$ have been individually fitted 
  to constants plus $L^{-1/2}$ corrections.}
\label{fig1}
\vskip-2mm
\end{figure}

The standard finite-size scaling ansatz \cite{barber83} for a quantity $A$ that scales as $|T-T_c|^\kappa$ in the thermodynamic limit is
(neglecting corrections from irrelevant fields)
\begin{equation}
A(T,L)=L^{-\kappa/\nu}f[(T-T_c)L^{1/\nu}],
\end{equation}
where the exponent $\nu$ governs the correlation length, $\xi\sim |T-T_c|^{-\nu}$, and 
the scaling function $f(x)$ must have the form $x^\kappa$ for $x\to \infty$ to ensure the correct thermodynamic limit. With this form,
the singular behavior in a system of finite length is cut off at $\xi \sim L$, i.e., at $|T-T_c| \sim L^{-1/\nu}$.  In the 2D $J=\pm 1$ spin
glass, $T_c=0$ and a low size-dependent energy scale was identified in previous works, $T^*(L) \sim L^{-\Theta_S}$, where $\Theta_S \approx 0.50$ is an 
exponent quantifying the entropy due to zero-energy clusters; flipping a cluster of linear size $l$ reduces the entropy by
$\Delta S \sim {l}^{\Theta_S}$ \cite{saul93,thomas11}. The finite-size scaling relation then changes to
\begin{equation}
A(T,L)=L^{-\kappa \Theta_S}f(TL^{\Theta_S}).
\label{alt}
\end{equation}
Thomas {\it et al.}~showed that the specific heat exponent is $\alpha=1-2/\Theta_S$ \cite{thomas11}. Then, at $T=0$, Eq.~(\ref{alt}) with 
$\kappa=\alpha$ predicts that the finite-size energy correction (per spin) should be $\Delta E_0 = E_0(L)-E_0(\infty) \sim L^{-2}$. This 
form was obtained based on a different scenario in Ref.~\onlinecite{campbell04} and was consistent with data for periodic systems. In our PT
simulations we generated a much larger number of samples for all system sizes up to $L=32$, to obtain a more reliable estimate of the $L^{-2}$
correction. As shown in Fig.~\ref{fig1}(a), the agreement with the prediction is excellent. The extrapolated infinite-size energy based on system
sizes for which no further scaling corrections are statistically important is $E_0=-1.40192(2)$, where the number within parentheses 
indicates the one-standard-deviation statistical error. This value is in good agreement with the best previous result, $E_0=-1.401938(2)$,
from open-boundary systems \cite{palmer99} (see also Ref.~\cite{campbell04}). Using the more precise value to constrain the fit we obtain
$\Delta E_0 = aL^{-2}$ with $a=1.230(2)$. 

We evaluate $\langle q^2\rangle$ according to Eq.~(\ref{ea})
and extrapolate it to infinite size, as shown in Fig.~\ref{fig1}(b). Since long-range order is expected
at  $T=0$, the exponent $\kappa=0$ in Eq.~(\ref{alt}) and the size dependence reflects a correction of the form 
\begin{equation}
\langle q^2(L)\rangle - \langle q^2(\infty)\rangle \propto L^{-\Theta_s},
\label{q2corr}
\end{equation}
derived in Ref.~\cite{thomas11}. With data for $6 \le L \le 24$ our independent estimate of the 
exponent is $\Theta_S = 0.60(3)$ for even $L$ and $\Theta_S = 0.52(3)$ for odd $L$. Fixing $\Theta_S=1/2$, as was also done in the data
analysis in Ref.~\cite{thomas11}, fits for both even and odd sizes are good and mutually consistent for $L \ge 9$. The extrapolated order parameter
is then $\langle q^2(\infty)\rangle = 0.373(3)$, which is roughly consistent with the previous estimate, $\langle q^2(\infty)\rangle = 0.395(10)$,
from large systems at low but non-zero temperature
\cite{thomas11}.

\section{Kibble-Zurek scaling}
\label{sec:kz}

Turning to the SA simulations, we measure the time $t$ in the standard way in units of MC sweeps, where each sweep consists of 
$N$ attempted Metropolis flips of randomly selected spins. Starting in equilibrium at $T_{\rm ini}=8$, we anneal to $T=0$ in
$t_{\rm max}$ MC sweeps according to the power-law protocol in Eq.~(\ref{tprotocol}).
We define the velocity for $r=1$ as $v=T_{\rm ini}/t_{\rm max}$ and use this definition of $v$ as the inverse of the total annealing time also
for $r=2$ and $4$. We collect expectation values at the final temperature $T=0$, using millions of samples for smaller sizes and several thousand 
for the largest systems. There is significant self-averaging and the error bars are small even for the largest sizes (much smaller 
than the plot symbols in the graphs below).

In standard KZ scaling, for a process stopping at the critical point, a singular quantity $A$ depends on the velocity and the 
system size according to the form \cite{zhong05,chandran12,liu14}
\begin{equation}
A(v,L)=L^{-\kappa/\nu}g(v/v_{\rm KZ}),
\label{avl}
\end{equation}
where the ``critical'' KZ velocity,
\begin{equation}
v_{\rm KZ} \propto L^{-\bar z-1/(\nu r)},
\label{vkzdef}
\end{equation}
is the velocity separating fast and slow processes, $\bar z$ is the dynamic exponent, and $g \to {\rm constant}$ when $v \to 0$.
Variants of this ansatz have been confirmed in uniform systems \cite{zhong05,chandran12,liu14} as well as in the 3D Ising spin glass
(where $T_c > 0$) \cite{liu15}. It has proved to be a reliable way to extract the dynamic exponent, especially if $T_c$ and $\nu$ are known,
e.g., for the 3D Ising glass $z \approx 6.0$ was obtained using KZ scaling \cite{liu15} and this value is in excellent agreement with a recent
result from a completely different apprroach \cite{fernandez16}. If $\nu$ is not known, it can be obtained along with $\bar z$ by combining results for
different $r$ values in the protocol (\ref{tprotocol}) \cite{liu14}, and in principle $T_c$ can be determined by using an extended scaling
ansatz \cite{zhong05,liu14}. Note that $v_{\rm KZ}$ in Eq.~(\ref{vkzdef}) is only determined up to an essentially arbitrary factor that can
be fixed by using some criterion once the scaling function $g(v/v_{\rm KZ})$ in Eq.~(\ref{avl}) has been determined, e.g., based on some
small deviation from the saturated equilibrium value.

The dynamic exponent relates the relaxation time scale $\tau$ to the equilibrium correlation length; $\tau \sim \xi^{\bar z}$. For given velocity, 
in the thermodynamic limit the correlation length saturates at
\begin{equation}
\xi_v \sim v^{-1/(\bar z+1/(\nu r))},
\label{xiv}
\end{equation}
and for finite system size the saturation velocity scales as $\xi_v \sim L$, i.e., $v \sim L^{-\bar z-1/(\nu r)}$ demarks the ``freezing'' of the
system. However, this analysis has neglected the entropic scale $L^{-\Theta_S}$ present in the $J=\pm 1$ spin glass model in equilibrium. This
new scale should also carry over to velocity scaling. Expressed as a length scale, the entropic scale is, $\xi_S \sim \xi^{1/(\nu \Theta_S)}$,
where presumably $\nu \approx 3.6$ \cite{katzgraber04,katzgraber07,fernandez16,toldin11} as in the model with normal-distributed couplings. 
In analogy with the equilibrium finite-size scaling behavior in the presence of the entropic scale \cite{thomas11}, since $\nu \Theta_S > 1$ and
$\xi_S \ll \xi$ the quasi-static behavior should set in for SA when $\xi_S \sim \xi_v^{1/(\nu \Theta_S)} \sim L$, which together with
Eq.~(\ref{xiv}) gives the entropy-driven analog of the KZ velocity
\begin{equation}
v_S \propto L^{-(\bar z v\Theta_S + \Theta_S/r)}.
\label{vs}
\end{equation}
We can define a more practical dynamic exponent for finite-size scaling purposes as
\begin{equation}
z=\bar z\nu\Theta_S,
\end{equation}
which gives the critical quasi-static velocity
\begin{equation}
v_S(L) \sim L^{-z-\Theta_S/r}
\end{equation}
in the same form as the original KZ velocity (\ref{vkzdef}) with $\nu$ replaced by $1/\Theta_S$, and the following modified KZ finite-size
scaling form:
\begin{equation}
A(v,L)=L^{-\kappa \Theta_S}g(v L^{z+\Theta_S/r}).
\label{avt2}
\end{equation}
We will test this hypothesis with SA data in the following sections.

\subsection{Order parameter}

\begin{figure}[t]
\center{\includegraphics[width=7.5cm, clip]{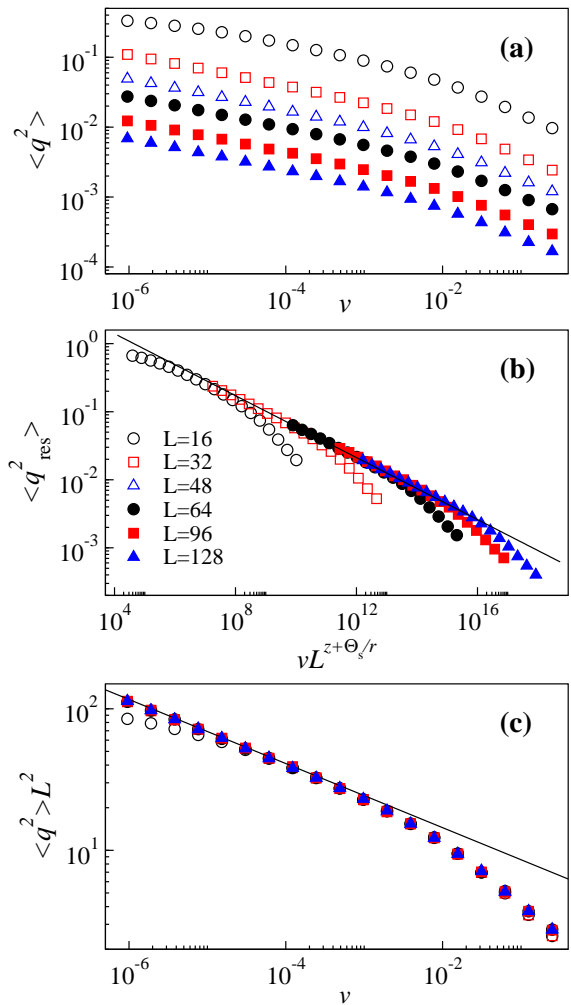}}
\vskip-1mm
\caption{(a) EA order parameter squared vs the velocity in linear ($r=1$) SA runs for different system sizes. (b)  Velocity scaling, where
  $\langle q^2\rangle$ has been rescaled by the size correction, $\langle q^2_{\rm res}\rangle = \langle q^2\rangle/(a+bL^{-1/2})$ with $a$ and $b$
  the constants of the even-$L$ fit in Fig.~\ref{fig1}(b), and the horizontal axis has been scaled with the optimal exponent  $z+\Theta_S=8.83(4)$
  (obtained with a data-collapse procedure using many system sizes between $L=32$ and $128$). The line has the expected slope $-2/(z+\Theta_S)$
 in the power-law scaling regime. (c) The same data scaled according to the third line of Eq.~(\ref{eq:q_scaling}) along with a line with the
 same slope as in (b).}
\label{fig2}
\vskip-1mm
\end{figure}

Results for the EA order parameter in linear SA runs are shown in Fig.~\ref{fig2}(a). For fixed velocity, the squared order parameter
$\langle q^2\rangle$ drops rapidly with increasing system size. In Eq.~(\ref{avt2}) we have $\kappa=0$ as in the equilibrium scaling of this quantity,
but since the correction to the asymptotic value of $\langle q^2\rangle$ is large, as seen in Fig.~\ref{fig1}, we first divide it out based
on the form in Eq.~(\ref{q2corr}). The so rescaled order parameter is, thus,
\begin{equation}
 \langle q^2_{\rm res}\rangle = \frac{ \langle q^2\rangle }{a+bL^{-\Theta_S}},
\label{qres}
\end{equation}
where we use $\Theta_S=1/2$ and the constants $a$ and $b$ from the fit in Fig.~\ref{fig1}(b). With this definition $\langle q^2_{\rm res}\rangle \to 1$ for
$v \to 0$ for all system sizes (up to small deviations due to inaccuracies of the fitted parameters and neglected higher-order corrections).
As shown in Fig.~\ref{fig2}(b), we then rescale the velocity by the size-dependent KZ velocity $L^{-z-\Theta_S}$, optimizing the value of the
exponent $z+\Theta_s$ for the best data collapse for large systems and low velocities. The data-collapse procedure is discussed further
in Appendix \ref{appendixa}. Here we just note that the goodness of the data collapse is quantified by a fit of data for all included
system sizes to a flexible function representing the scaling function $g$ in Eq.~(\ref{avt2}).

The KZ scaling form (\ref{avt2}) discussed above only applies for sufficiently low velocity and the inability to collapse the data at high velocities
in Fig.~\ref{fig2} is not surprising. We observe power-law behavior over a wide range of scaled velocities and also see a flattening-out toward the expected
constant behavior on the low-velocity side (which we can see more clearly for smaller system sizes, as discussed in detail in Sec.~\ref{sec:small}).

According to the general non-equilibrium finite-size scaling form discussed in Ref.~\cite{liu14}, adapted to the present case where $1/\nu$ is replaced
by $\Theta_S$, we expect that the squared order parameter can be written in the following way in three distinct velocity regimes:
\begin{equation}
 \langle q^2\rangle \propto \begin{cases}
     \sum\limits_{n} c_n(vL^{z+\Theta_s/r})^n,~~   v \alt v_{\rm KZ},\\
     \\
     (vL^{z+\Theta_s/r})^{-x} =  L^{-2}({1}/{v})^x,~~  v_{\rm KZ} \alt v \alt 1,\\
     \\
     L^{-2}\sum\limits_{n} c_n(1/v)^n,~~  v \agt 1.\\
     \end{cases}
\label{eq:q_scaling}
\end{equation}
Here we think of $v_{\rm KZ}$ as the velocity separating the near-equilibrium and power-law scaling behaviors.  The factor $L^{-2}=N^{-1}$ 
on the second and third line represents the overall size dependence in the limit where $\xi_v \ll L$. In order for the two expressions on the middle
line to be equal, the exponent $x$ has to be given by
\begin{equation}
x=\frac{2}{z+\Theta_s/r}.
\label{x_exponent}
\end{equation}
The Taylor-expandable near-equilibrium behavior on the first line of Eq.~(\ref{eq:q_scaling}) should smoothly connect
to the first power-law form on the second line, through a cross-over region in the scaling function (\ref{avt2}). In the high-velocity limit, the third case
above, the behavior can be expressed as a series in $1/v$, and this series has to be smoothly connected to the form $L^{-2}v^{-x}$ on the second line.

For convenience we denote the often occurring generalized KZ exponent by $\sigma(r)$,
\begin{equation}
\sigma(r)=z+{\Theta_S}{r^{-1}}.
\label{a}
\end{equation}
One can use Eq.~(\ref{eq:q_scaling}) for given $r$ to extract this exponent either from the high-velocity side, by fitting a straight line to
$\ln\langle q^2_{\rm res}\rangle$ versus $\ln(1/v)$, the slope of this line being the exponent $x=2/\sigma(r)$ in Eq.~(\ref{x_exponent}), or by adjusting
$\sigma(r)$ so that $\langle q^2_{\rm res}\rangle$ versus $vL^{\sigma(r)}$ in the power-law and equilibrium regimes collapse onto a common scaling function
for different $L$. These two methods were also illustrated in Figs.~\ref{fig2}(b,c). To cancel out the leading equilibrium finite-size corrections,
in the low-velocity analysis in panel (b) we used the rescaled order parameter, while in the high-velocity analysis in panel (c) the original
data were used.

The analysis from the high-velocity side, Fig.~\ref{fig2}(c), can include data only in the strict power-law regime,
unless high-velocity corrections are taken into account. The behavior as $v \to \infty$ is clearly non-universal, with the curve tending to the
equilibrium value at the initial temperature. The data-collapse method in Fig.~\ref{fig2}(b) potentially can lead to better statistical precision
on the extracted exponent if a substantial amount of data is available in the low-velocity cross-over and equilibrium regimes, where the power-law scaling 
no longer holds. Due to the slow dynamics of the Ising glass model, reflected in the large value of the KZ exponent, $z + \Theta_S \approx 9$, we can only
reach the equilibrium and cross-over regions clearly for very small system sizes, which we discuss further in Sec.~\ref{sec:small}. In Fig.~\ref{fig2}(b),
in order to minimize finite-size corrections (beyond those appearing explicitly in the KZ form), we exclude the smallest systems, and, therefore, mainly
collapse data in the power-law region (though some of the included low-$v$ data do deviate from the pure lower-law). By systematically monitoring the
goodness of the fit as small systems are gradually excluded, we find that the exponent settles to $z+\Theta_S=8.83(4)$ when the fit becomes
statistically sound ($\chi^2$ per degree of freedom is close to one).

\begin{figure}[t]
\center{\includegraphics[width=7.5cm, clip]{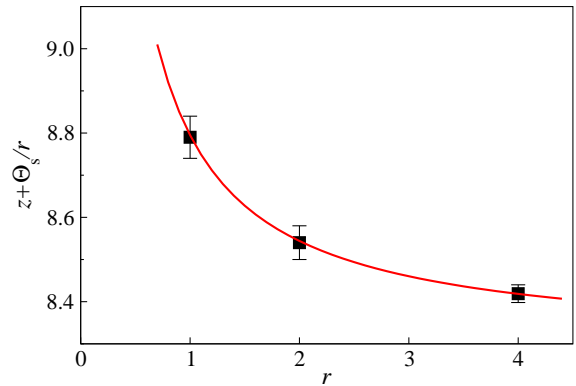}}
\vskip-1mm
\caption{The KZ scaling exponent vs the exponent $r$ in the SA protocol (\ref{tprotocol}),  along with a fit giving $z=8.28(3)$ and $\Theta_S=0.55(6)$.
The $r=1$ point was obtained using the data analysis in Fig.~\ref{fig2}(b) and similar data for $r=2$ and $r=4$ are presented in Appendix \ref{appendixb}.}
\label{exponent}
\vskip-1mm
\end{figure}

To disentangle Eq.~(\ref{a}) and obtain the exponents $z$ and $\Theta_S$, it is in principle sufficient to work with two different values of $r$ in the
annealing protocol, Eq.(\ref{tprotocol}), and extract $\sigma(r_1)$ and $\sigma(r_2)$. Here, as a further consistency check we use three different values,
$r=1,2,4$, and fit the resulting $\sigma(r)$ to the expected form (\ref{a}) with $z$ and $\Theta_S$ optimized for the best fit. The procedure is illustrated
in Fig.~\ref{exponent}. The $r=2$ and $r=4$ data sets corresponding to Fig.~\ref{fig2}(b) for $r=1$ are presented in Appendix \ref{appendixb}.
The $\sigma(r)$ data points are completely consistent with the expected $r$-dependence in Eq.~(\ref{a}), and a fit delivers the exponent values $z=8.28(3)$
and $\Theta_S=0.55(6)$. Fixing $\Theta_S=1/2$  does not significantly alter the estimate of $z$.

\subsection{Mean energy}

\begin{figure}[t]
\center{\includegraphics[width=8cm, clip]{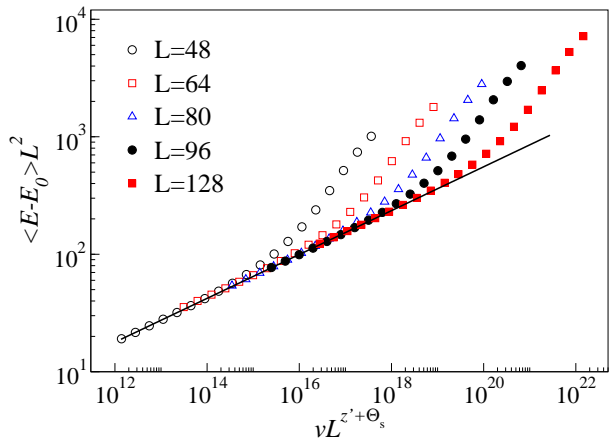}}
\vskip-2mm
\caption{Mean energy above the infinite-size ground state energy $E_0$ in $r=1$ SA runs.
The results have been divided by the $L^{-2}$ equilibrium size-dependence and the horizontal axis was rescaled according to the generalized
KZ hypothesis (\ref{avt2}), with $\kappa\Theta_S=2$ and the optimum-collapse value $z'+\Theta_S=10.80(8)$ of the scaling exponent. 
The line shows the expected slope $2/(z'+\Theta_S)$ in the power-law regime.}
\label{fig3}
\vskip-2mm
\end{figure}

Forms analogous to Eq.~(\ref{eq:q_scaling}) for the order parameter hold for other singular quantities as well.  On the left-hand side 
the critical size-dependence in equilibrium, i.e., the factor $L^{-\kappa\Theta_S}$ in Eq.~(\ref{avt2}), should be divided out
(and in principle finite-size corrections can also be divided out, as we did above for $\langle q^2\rangle$). The factor $L^{-2}$ on the second and third
lines should be replaced by $L^{\kappa\Theta_S-2}$. To study the singular part of the
energy, we first subtract the infinite-size value $E_0$ from the velocity dependent energy $E(v,L)$ and use $\kappa\Theta_S=(|\alpha|+1)\Theta_S=2$
in Eq.~(\ref{avt2}). We again optimize the data collapse with small systems and high velocities excluded. Fig.~\ref{fig3} shows $r=1$ results and
similar $r=2,4$ plots are presented in Appendix \ref{appendixb}.

Combining the results for $z'+\Theta_S/r$ for the different $r$ values, we can again, as in Fig.~\ref{fig2}(c), disentangle the exponents. Interestingly,
here we obtain a clearly different dynamic exponent, $z'=10.32(7)$, than the previously extracted exponent $z=8.28(3)$ governing the EA order parameter,
while $\Theta_S=0.5(1)$ is consistent with the previous value. Fixing $\Theta_S=1/2$ we can reduce the error bar on the dynamic exponent; $z'=10.31(4)$.

\subsection{Scaling results for small system sizes}
\label{sec:small}

In the previous sections we discussed velocity scaling for systems sufficiently large for no significant subleading
finite-size scaling corrections to remain (to within the statistical precision of the data). For these system sizes we can reach
well into the power-law scaling regime (the linear part of the scaling function graphed on a log-log scale), but not very far into the
cross-over into the regime where the systems approach and reach equilibrium, i.e., corresponding to the first line in Eq.~(\ref{eq:q_scaling}).
It is important to test the scaling behavior also here, to make sure that the final relaxation stage is governed by the same dynamic exponent
as the power-law regime. Because of the large dynamic exponents, we are in practice limited to small system sizes in this velocity regime.
We show here that useful results further supporting the dual time-scale picture can still be obtained.

\begin{figure}[tp]
\center{\includegraphics[width=7.5cm, clip]{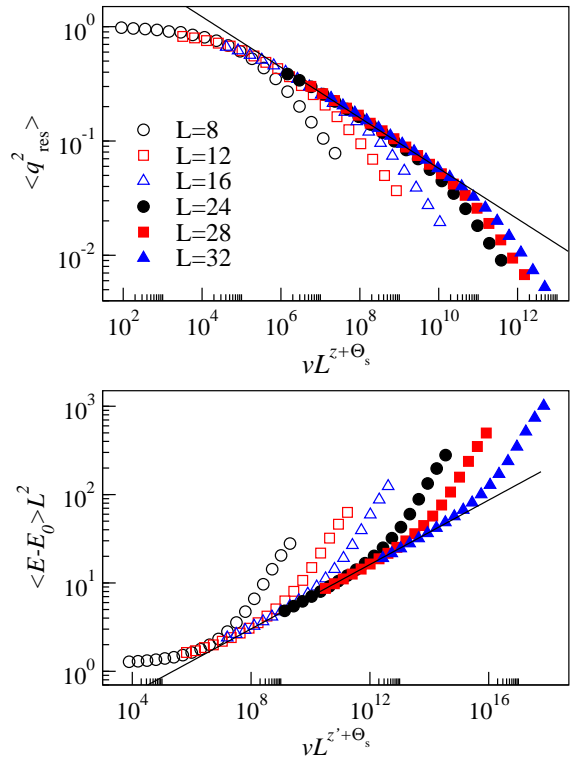}}
\vskip-2mm
\caption{Generalized KZ scaling analysis of the kind presented in Figs.~\ref{fig2} and \ref{fig3} but for smaller systems for which the cross-over
  toward equilibrium can be more clearly observed. The velocity scaling exponents are $\sigma=z+\Theta_s=9.01$ and $\sigma'=z'+\Theta_S=10.8$ for
  the EA order parameter (upper panel) and the excess energy (lower panel), respectively. The lines have slopes $2/\sigma$ and $2/\sigma'$,
corresponding to the expected exponents in the power-law scaling regime.}
\label{qesmall}
\vskip-2mm
\end{figure}

Figure \ref{qesmall} shows $r=1$ results for lattice sizes in the range $L=8$ to $32$, with the exponent $\sigma=z+\Theta_S$ adjusted for
best overall data collapse. Here the data-collapse procedure included all the system sizes shown, again excluding high velocities
where no data collapse can be expected. In the low-velocity limit the rescaled order parameter approaches $1$, while the scaled energy
tends to the value of the constant $a \approx 1.23$ extracted in Fig.~\ref{fig1}(a).
We obtain the exponents $\sigma=9.01(5)$ and $\sigma'=10.9(2)$ for the EA order parameter and the
energy, respectively. These values are very close to those obtained for larger system sizes in Figs.~\ref{fig2} and \ref{fig3}, $\sigma=8.83(4)$ and
$\sigma'=10.80(8)$, respectively, demonstrating the stability of the results. We did not include the smallest systems in the previous analysis
because, although the exponent values do not differ much, we can not obtain a statistically fully satisfactory value of the goodness of the fit ($\chi^2$
per degree of freedom close to $1$) when all data are included in a common fit, given the small error bars of the SA data and small but statistically
significant effects of neglected finite-size scaling corrections for the smaller systems.

Given the good agreement we have demonstrated between different system sizes and velocity regimes, we judge that the significant difference between
the dynamic exponent for the excess energy and the EA order parameter, $z'-z \approx 2$, cannot be explained by neglected scaling corrections. The
dual time scales are therefore a real aspect of the relaxation of the 2D $J = \pm 1$ spin glass.

\subsection{Minimum energy}

When applying SA to an optimization problem, it is in general better to keep track of the minimum energy (cost function) $E_{\rm min}$ reached during
an entire SA run, instead of computing the mean energy or only using the energy at the end of the run. Even in very slow annealings the minimum energy is
occasionally lower than the energy after the final MC step at $T=0$. Therefore, the disorder-averaged $\langle E_{\rm min}\rangle$ should be lower than
$\langle E\rangle$. An important question then is whether the scaling of the two quantities is the same or not. We address this question next.

\begin{figure}
\center{\includegraphics[width=8cm, clip]{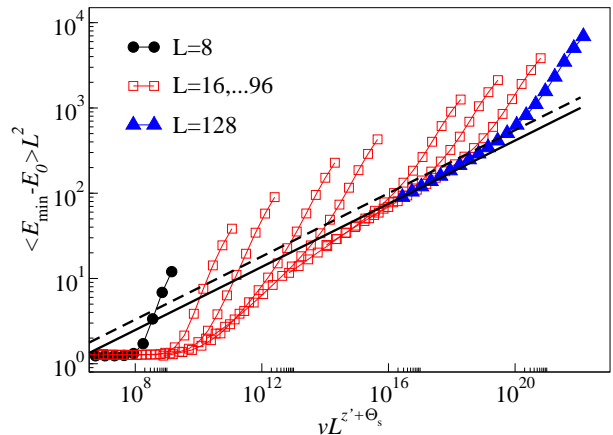}}
\vskip-2mm
\caption{Scaling of the minimum energy reached during any time in $r=1$ SA runs for $L=8,16,24,32,48,64,72,96,128$.
  The exponents used for the rescaling of both the axes
  are the same as those in Fig.~\ref{fig3}. The solid line has the expected slope $2/(z'+\Theta_S)$ and is drawn in close proximity
  to the data for the largest system sizes. The dashed line has the same parameters as the line in Fig.~\ref{fig3}.}
\label{emin}
\vskip0mm
\end{figure}

For each SA run, we save the minimum energy in any of the 64 replicas running in parallel and average over samples.
Figure \ref{emin} shows results for $r=1$, scaled using the same exponents as in Fig.~\ref{fig3}. The scaling collapse is very good also here,
and the optimized scaling exponent for this case is also statistically equal to the one obtained before. Overall the minimum energy
values are, as expected, below those for the mean energy. With the range of system sizes used here we can see the full equilibrium behavior (the
flat portion, where the value corresponds to the prefactor of the $L^{-2}$ correction in Fig.~\ref{fig1}) as well as the cross-over
into the power-law scaling regime. In the graph we also draw a straight line with exactly the same parameters as the line drawn through the power-law
scaling portion of the collapsed mean energy data in Fig.~\ref{fig3}. In $\langle E_{\rm min} - E_0\rangle L^2$ we observe that larger systems are needed to
observe the same slope---we see that the scaling function (onto which the data collapse) exhibits some curvature. Nevertheless, with increasing size
the functional form appears to approach a line with the same slope as before. It is possible that the curves for $L\to \infty$ actually approach exactly the
same line (not just the same slope but also the same constant) as the one for $\langle E - E_0\rangle L^2$ in Fig.~\ref{fig3}. If so, the asymptotic power-law
scaling of the two quantities would be exactly the same, and the advantages (in optimization applications) of monitoring $E_{\rm min}$ instead of $E$
would only appear as the behavior crosses over toward the equilibrium behavior.

\begin{figure}
\center{\includegraphics[width=8cm, clip]{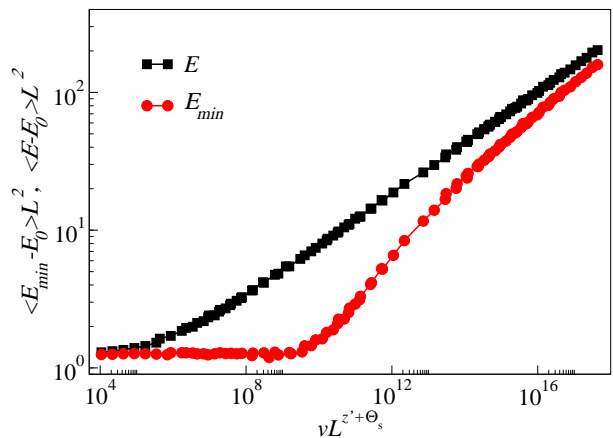}}
\vskip0mm
\caption{Scaling function for the mean energy and the mean lowest energy in $r=1$ SA runs. Data points $(v,L)$ for system sizes
  in the range $L=8$ to $128$ were used, for each size excluding $v$ points that deviate from the common data collapsed form.
  The value of the scaling exponent is the same as in Fig.~\ref{fig3}: $z'+\Theta_s=10.8$.}
\label{function}
\vskip-2mm
\end{figure}

We conclude that the minimum energy collected during SA runs converges to the ground state energy on the same time scale $L^{z+\Theta_s/r}$ as
the convergence of the mean energy. The scaling functions are different, reflecting an overall lower value of the minimum energy than the
mean energy for given scaled velocity $vL^{z'+\Theta_S/r}$. In Fig.~\ref{function} we show the two $r=1$ scaling functions in the same graph.
We have combined data from other figures but only included points that fall very close to the common scaling functions. Here one
can read off that $E_{\rm min}$ ultimately converges (the curve flattens out to a constant) about $10^4$ times faster than $E$. This factor
depends on the details of how $E_{\rm min}$ is computed in the simulations. In our case, we carried out 64 simulations in parallel for each
coupling sample and monitored the lowest energy in any of these simulations. Clearly, upon increasing the number of parallel runs
$E_{\rm min}$ will converge faster, thus pushing the scaling function further to the right.

\section{Discussion and Conclusions}
\label{sec:disc}

The existence of two different dynamic exponents at first sight appears to contradict the standard picture of critical dynamics, where the slowest 
mode is associated with the fluctuation of the order parameter. The coupling of the energy to the order parameter (via defects) normally implies
that the asymptotic energy auto-correlations are also determined by $z$. Thus, in the standard  scenario, there is a single exponent governing the dynamic
scaling of all quantities, except ones that are explicitly constructed to only sense {\it faster} modes. 

Given the unusual behavior, it is natural to ask whether scaling corrections may explain the rather large difference between the two dynamic exponents,
$z'-z \approx 2$, so that there would actually only be a single common exponent for the energy and order-parameter relaxation. The fact that both small 
and large systems lead to the same exponents (as shown in Figs.~\ref{fig2} \ref{fig3}, and \ref{qesmall}) speaks against the existence of large finite-size 
corrections beyond the leading corrections that we have included (based on the analysis of equilibrium results in Fig.~\ref{fig1}). 
Since the two groups of system sizes also probe regimes closer to (the smaller sizes) and further away from (the larger sizes) equilibrium,
the good agreement between the exponents also indicates that any velocity corrections must be small. The insignificance of velocity corrections in the
power-law regime is also supported by the fact that scaling (data collapse) works extremely well over 1-2 orders of magnitude of the scaled energy and 
order parameter for the larger system sizes, with no deviations detected from the power-law behavior. In this regard the behavior is similar to that in the 
3D Ising spin glass, for which also no corrections to velocity scaling were needed to collapse data analyzed within the KZ framework \cite{liu15}. Subsequently, 
the value of the dynamic exponent extracted was reproduced with a completely different approach \cite{fernandez16}. 

\begin{figure}[t]
\center{\includegraphics[width=7cm, clip]{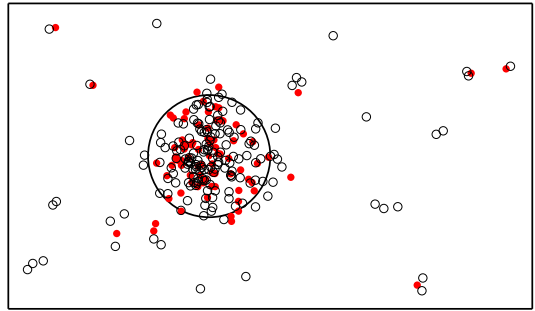}}
\vskip-1mm
\caption{Conceptual illustration of the essential features of the energy landscape of the $J=\pm 1$ model, as suggested by droplet theory and
  our study. Ground states (red solid circles) are clustered and typical states fall within a small region (large circle) of the configuration
  space. Low-energy excitations (black open circles) are predominantly located in the same region.}
\label{fig4}
\vskip-2mm
\end{figure}

We also point out that the fact that the critical temperature $T_c=0$ is known exactly removes one of the potential flaws in data-collapse approaches, 
namely, that the agreement between the scaled data sets may be artificially improved by the procedure of adjusting exponents as well as the 
critical-point value, thus leading to systematical errors in all the fitting parameters. In the present case we only adjusted a single exponent 
$\sigma(r)=z+\Theta_S/r$ (and similarly with $z \to z'$) independently for each of three annealing protocols (exponent $r=1,2,4$ in the power-law
annealing form), and when combining the results according to the proposed generalized KZ scaling form,
Eq.~(\ref{avt2}), the entropy exponent $\Theta_S$ comes out very close to its previously calculated value. It is hard to believe that 
this success in reproducing a non-trivial thermodynamic exponent in a dynamical approach could be a mere coincidence. Another consistency check
is provided by the scaling of the mean energy and the lowest energy, shown in Fig.~\ref{function}. Their scaling functions are very different,
with the lowest energy converging much faster to the equilibrium value, yet the extracted dynamic exponents are the same for both of them.
Thus, there are many reasons 
to trust the exponents extracted here, as well as their error bars (which we have computed based on extensive bootstrapping and considering different 
windows of velocities and system sizes). We conclude that the difference $z'-z \approx 2$ is too large to be explained by scaling corrections, unless 
the flow of the exponents to their true values is so slow that the changes cannot be detected at all in the size and velocity regimes considered here. 
Such an extremely slow convergence is unlikely and would in itself be remarkable and beyond current understanding of out-of-equilibrium scaling.

We next provide a physical explanation for our findings. 
The dual dynamic scales should be related to the phenomenon of droplet entropy stabilizing the EA order parameter of the 2D $J=\pm 1$ Ising glass  
when $T \to 0$ in equilibrium. The backbone of the spin-glass cluster has a fractal dimension $d_{\rm f} < 2$ \cite{hartmann08} and, thus, does not 
represent long-range order on its own \cite{thomas11}. The ground states are strongly clustered within a small region (and its spin-reflected
counterpart), which implies that these states are related to each other by flipping small (compared to the system size) droplets; flips of large 
droplets throw the system into atypical regions that are statistically insignificant in the thermodynamic limit. Although the absence of
order at $T>0$ implies that low-energy excitations must be spread out over a large region of the configuration space, the ground state region
should also have a much higher density of low-energy states than other regions (since these states can be obtained from the ground states by 
flipping small clusters). Thus, there should exist a region of typical ground states and low-energy states, illustrated in Fig.~\ref{fig4}, and the large 
entropy drives the system toward this region under annealing. In the typical region, the EA order parameter, Eq.~(\ref{ea}), has essentially the 
same distribution for replicas in low-energy states as for those strictly in ground states, and, therefore, the order parameter can converge 
even when a significant fraction of the replicas remain in excited states. Our scaling results show that the final relaxation of the system 
involves transitions between excited states into ground states located in the same high-density region, and that the time scale for this is 
significantly longer (approximately by a factor $L^2$) than that for reaching the high-density region.

\begin{figure}[t]
\center{\includegraphics[width=5.5cm, angle=-90, clip]{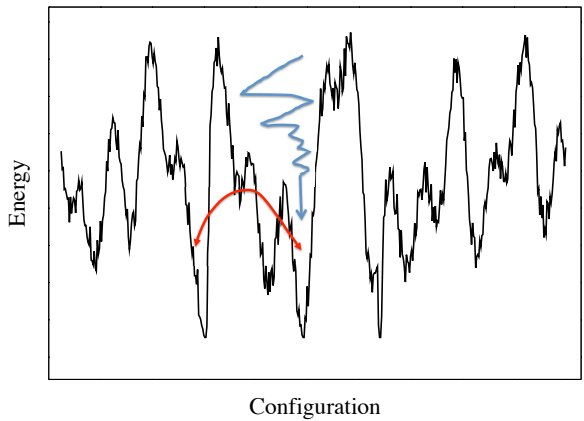}}
\vskip-1mm
\caption{Conceptual one-dimensiomal analog of our proposal for the difference between equilibrium and non-equilibrium dynamics of the spin glass. In the
  equilibrium at low temperatures, the slowest dynamic time scale corresponds to fluctuations between regions surrounding different low-energy states. In a
  non-equilibrium SA simulation, at some point the system locks into one such funnel. The dynamic order-parameter exponent $z$ characterizes the process of
  reaching a funnel, while the energy exponent $z'$ governs the eventual relaxation to the bottom of the funnel.}
\label{e_config}
\vskip-2mm
\end{figure}

It should be noted that the $T \to 0$ relaxation dynamics we study here is different from the equilibrium dynamics at fixed temperature, with the
same kind of MC updates (here using the standard single-spin Metropolis algorithm). In previous works \cite{katzgraber05} it has been shown that
the equilibrium dynamic exponent $z_{\rm eq}$ depends on the temperature and $z_{\rm eq} \to \infty$ as $T \to 0$. This behavior is consistent
with the fact that the single-spin Metropolis algorithm is not ergodic at $T=0$---while some spins can be flipped without changing the
energy, not every ground state can be reached in this way. The equilibrium autocorrelation function at $T>0$ quantifies the way in which a
simulation explores the global configuration space, which at low temperatures corresponds to migrating between regions of states surrounding
different ground states. In contrast, in an SA simulation the system can be expected to become trapped in one of these regions---a ``funnel''
in the energy landscape, and when that happens the final relaxation corresponds to reaching the bottom of the funnel. The different kinds of
dynamical processes are illustrated in Fig.~\ref{e_config}.

We have argued above that the dynamic exponent $z$ characterizes the time scale upon which
the system reaches the region of the configuration space with a large number of low-energy states, i.e., the funnels. In our picture,
the larger energy exponent $z'$ then characterizes the time scale of trapping of the system in local energy minimas along the
``walls'' of the funnel, and the fact that we observe power-law scaling implies that the barriers (in energy and entropy) do not
grow sufficiently large with increasing system size to cause an exponential slowing down. In principle, there could also be funnels
with a lowest energy larger than the ground state energy, but the fact that $z'$ is finite shows that such funnels must
have a statistically negligible weight, or are separated from ground-state funnels by barriers that grow only very slowly with the
system size (to maintain power-law scaling of the relaxation time).

Dynamic scaling is also interesting in the context of optimization. It has recently been argued that the best measure of optimization is not 
necessarily just the energy (the standard cost function), but the stability of the solution is also important and should be enhanced if the 
solution belongs to a dense region of similar solutions \cite{baldassi15,baldassi16}. A method was presented to enhance the ability to 
reach such regions, by using coupled replicas of the system. The 2D $J=\pm 1$ Ising spin glass may be an extreme case of a system harboring
a dense region of low-energy states, and we have shown here that SA finds this region efficiently even without artificial replicating,
as evidenced by the entropy-driven order parameter converging in polynomial time and even faster than the energy. In optimization, one may be
willing to accept a slightly sub-optimal solution, as measured by the energy, for a solution in a dense region that can be found on a much shorter
time scale. Clustering of solutions is also important when discussing the efficiency of QA protocols, where the measure of success is also ambiguous
and solution stability may be a desirable feature. QA of systems with discrete coupling distributions may also be affected by dual time scales,
due to mechanisms similar to those discussed here.

It would clearly be interesting to also study the KZ dynamics of the model with normal-distributed couplings, which has a unique 
ground state and likely different dynamic scaling. KZ scaling of $T\to 0$ SA simulations can also be used in other systems that do not 
order at $T>0$. Stimulated by the present work, the procedures were already applied to a planar vertex model encoding a class of 
reversible computing problems \cite{chamon16}. 

\begin{acknowledgments}
  We would like to thank Anatoli Polkovnikov, Claudio Chamon, and David Huse for stimulating discussions. The research was
  supported by the NSF under Grant No.~DMR-1410126 and by Boston University's Undergraduate Research Opportunities Program. Computations were carried
  out on Boston University's Shared Computing Cluster.
\end{acknowledgments}

\appendix

\begin{figure}
\centerline{\includegraphics[width=7.5cm]{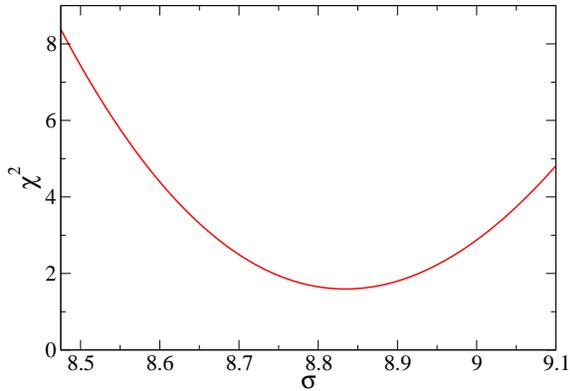}}
\vskip-2mm
\caption{Example for $r=1$ of the goodness of the linear fit of $\ln\langle q^2_{\rm res}\rangle$ versus $\ln(vL^\sigma_r)$ when the slope of the line
  is fixed at $-2/\sigma$. The data used here is shown as the middle set (triangles) in Fig.~\ref{fig:linear}. A scan is performed as a function
  of the exponent $\sigma=z+\Theta_s$ and the minimum of $\chi^2$ is identified for the optimum $\sigma$.} 
\label{fig:x2sigma}
\vskip-1mm
\end{figure}

\section{Data collapse procedures}
\label{appendixa}
        
Here we give further details on the data-collapse procedures.
We take $r=1$ as an example and in the following simply use $\sigma$ to denote the exponent $\sigma(r=1)=z+\Theta_S$. In Fig.~\ref{fig2}(a,b)
we already illustrated how data are collapsed by optimizing $\sigma$. To characterize the goodness of the data collapse, we fit a high-order
polynomial to a set of data points $\{\ln\langle q^2_{\rm res}\rangle,\ln(vL^\sigma)\}$ for different $v$ and $L$, sweeping over $\sigma$ on a dense grid
and locating the optimal value (minimum $\chi^2$ for the fit). If a satisfactory collapse, $\chi^2/N_{\rm dof} \approx 1$, cannot be achieved we systematically
eliminate small system sizes and/or high-velocity points until a statistically good fit is obtained. Typically tens of data points are left in the good fit.
To estimate error bars, we perform bootstrapping, repeating the fitting procedure with many bootstrap samples and computing the standard deviation
of the optimal $\sigma$.

Here, for illustration purposes and to demonstrate the stability of the exponents extracted in Fig.~\ref{fig2},
we discuss a slightly simpler method for analyzing only the power-law regime and including only the three or four lowest available
velocities for three system sizes; $L=72,80,96$. For these sizes, even at the lowest velocity that we have studied, $v=8/t_{\rm max}=2^{-17}$, the systems are far
from equilibrium but, as we will show, they fall within the power-law scaling regime described by the middle line in Eq.~(\ref{eq:q_scaling}). Graphing on a
log-log scale, we then expect all points to fall on a common line with slope $x$ given by Eq.~(\ref{x_exponent}) if the horizontal axis is appropriately rescaled
as $vL^\sigma$. We use the required $r=1$ line slope $-2/\sigma$ to constrain the fit to the form $-(2/\sigma)\ln(vL^\sigma)+b$, i.e., for given $\sigma$ in
the scaling procedure $b$ is the only adjustable parameter. We scan over a dense grid of $\sigma$ values, perform the constrained line fit for each case, and keep
track of $\chi^2$ to locate the minimum value; see Fig.~\ref{fig:x2sigma} for an illustration. The optimum $\sigma$ value is the result.

\begin{figure}[t]
\centerline{\includegraphics[width=7.5cm]{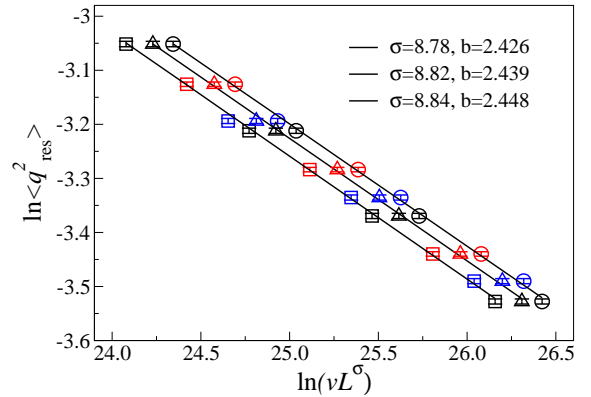}}
\vskip-2mm
\caption{Three different rescaled sets of data (indicated by different shapes of the graph symbols) obtained by bootstrap sampling of a large number
  of data bins for system sizes $L=72$ (black), $80$ (red), and $96$ (blue). The three lines are the best fits to the form
  $\ln\langle q^2_{\rm res}\rangle=-({2}/{\sigma})\ln(vL^\sigma)+b$, and the values of $\sigma,b$ shown for each case corresponds to a $\chi^2$-minimum
  such as the one in Fig.~\ref{fig:x2sigma}.}
\label{fig:linear}
\vskip-1mm
\end{figure}

\begin{figure}
\centerline{\includegraphics[width=7.25cm]{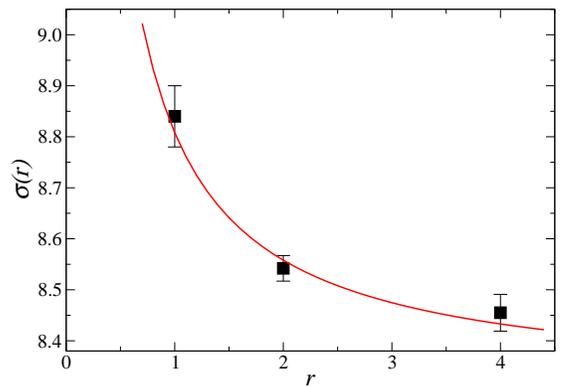}}
\vskip-2mm
\caption{The exponent $\sigma(r)$ for $r=1,2,4$ along with a fit to the form (\ref{a}). The individual exponents (fitting parameters) are $z=8.35(7)$,
and $\Theta_S=0.42(8)$, where the error bars were computed by repeated fits with Gaussian noise added [with standard deviation equal to the error bars on
$\sigma(r)$].}
\label{fig:a(r)}
\vskip-1mm
\end{figure}

Alternatively, according to the
second form of the middle line in Eq.~(\ref{eq:q_scaling}), we could also just consider $\ln(L^2 \langle q^2_{\rm res}\rangle)$ versus $\ln(1/v)$ and extract the
slope $x=2/\sigma$ (and again a good $\chi^2$ value would be an indication of being within the power-law scaling regime). The approach discussed here can, however,
also be generalized to include low-velocity data, where the power-law scaling no longer holds but the behavior is still described by the scaling function
$g(vL^\sigma)$, of which the power-law constitutes the limiting form for large $vL^\sigma$. This latter part can be fitted to a line, and points deviating from
it for smaller $vL^\sigma$ can be simultaneously fitted to a polynomial \cite{liu14}. Here we just consider the linear part, while in Fig.~\ref{fig2} we also
included lower-$v$ data but did not constrain the collapse by the line slope, instead obtaining the slope as a post-fit consistency check.

\begin{figure}
\center{\includegraphics[width=6cm, clip]{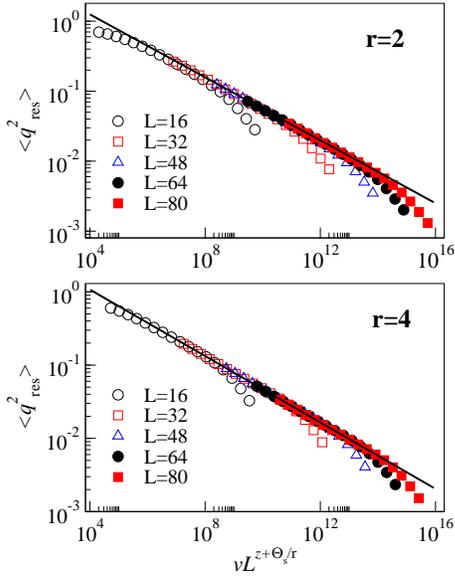}}
\vskip-2mm
\caption{Scaling collapse of the EA order parameter for $r=2$ and $r=4$. The scaling exponents are 
$\sigma(r=2)=8.57(3)$ and $\sigma(r=4)=8.42(2)$.}
\label{r2r4q}
\vskip-2mm
\end{figure}

\begin{figure}
\center{\includegraphics[width=6cm, clip]{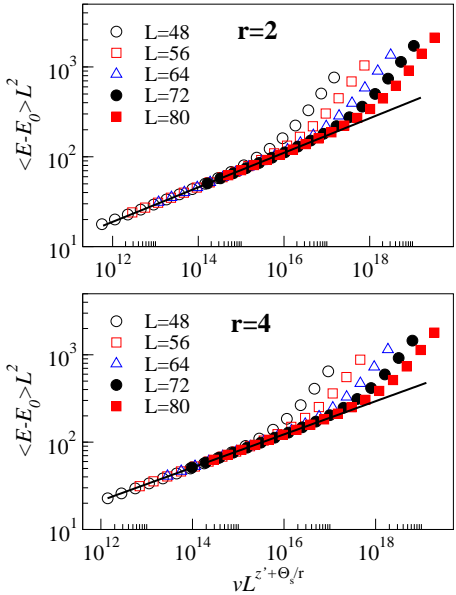}}
\vskip-2mm
\caption{Scaling collapse of the excess energy for $r=2$ and $r=4$. The scaling exponents are 
$\sigma'(r=2)=10.57(10)$ and $\sigma'(r=4)=10.44(6)$.}
\label{r2r4e}
\vskip-2mm
\end{figure}

Figure \ref{fig:linear} shows three different data sets along with the corresponding slope-constrained line fits.
The middle set of points is the original data set, while the left and right sets correspond to the extreme cases out of $200$ bootstrap samples. The standard
deviation of $\sigma$ computed from the bootstrap samples directly gives the error bar; in this case $\sigma=\sigma(1)=8.79(8)$. This value is completely
consistent with the value in the caption of Fig.~\ref{fig2}, but the error bar is larger because only the linear regime was used and the number of data
points is smaller.

Note that the same coupling realizations are used in SA runs with all velocities (where $v$ is if the form $2^{-n}$ for positive integers $n$), and the data
points for the same system size but different $v$ are therefore strongly correlated. The covariance predominantly corresponds to common up or down fluctuations of the value of the order parameter, and therefore the optimum
line slope, as extracted above, is not significantly affected, and it is not necessary to use the full covariance matrix in the fitting procedure.
The bootstrapping procedure properly account for the covariance since the same bins are randomly chosen for all velocities for a given $L$.

Using the same system sizes and velocities and repeating the same procedures for $r=2$ and $4$,
we obtain $\sigma(2)=8.52(6)$ and $\sigma(4)=8.48(7)$. Combining these results and
performing a fit to the expected $r$ dependence of $\sigma(r)$, Eq.~(\ref{a}), we obtain $z=8.35(7)$, and $\Theta_S=0.42(8)$, as shown in Fig.~\ref{fig:a(r)}.
These values are consistent with those presented in Sec.~\ref{sec:kz}, but again the error bars are larger due to the smaller amount of data used. We can then
conclude that the inclusion of also smaller sizes and lower velocities (including some data away from the power-law regime) in Fig.~\ref{fig2} did not
change the exponents to a noticeable degree relative to the case here, where only large system sizes far from equilibrium were used.

\section{Results for $r=2$ and $r=4$}
\label{appendixb}

For completeness we here present the data for $r=2$ and $r=4$, analyzed in the same way as the $r=1$ data in Figs.~\ref{fig2}(b) and \ref{fig3}.
Data-collapse plots for the EA order parameter and the excess energy are presented in Figs.~\ref{r2r4q} and \ref{r2r4e}, respectively. The exponent
values are given in the figure captions.

\end{document}